\documentclass{moriond}

\usepackage{graphicx}
\usepackage{amsmath,amssymb}
\usepackage{xspace}

\newcommand{\XeXe}{\mbox{Xe+Xe}\xspace}
\newcommand{\pp}{\mbox{$pp$}\xspace}
\newcommand{\PbPb}{\mbox{Pb+Pb}\xspace}
\newcommand{\sqn}{\mbox{$\sqrt{s_{\mathrm{NN}}}$}\xspace}
\newcommand{\sqs}{\mbox{$\sqrt{s}$}\xspace}
\newcommand{\Raa}{\mbox{$R_{\mathrm{AA}}$}\xspace}
\newcommand{\Taa}{\mbox{$T_{\mathrm{AA}}$}\xspace}
\newcommand{\avgTaa}{\mbox{$\langle \Taa \rangle$}\xspace}
\newcommand{\pT}{\mbox{$p_\mathrm{T}$}\xspace}

\newcommand{\Npart}{\mbox{$N_{\mathrm{part}}$}\xspace}
\newcommand{\Ncoll}{\mbox{$N_{\mathrm{coll}}$}\xspace}
\newcommand{\avgNpart}{\mbox{$\langle \Npart \rangle$}\xspace}
\newcommand{\avgNcoll}{\mbox{$\langle \Ncoll \rangle$}\xspace}
\newcommand{\akt}{\mbox{anti-$k_{t}$}\xspace}
\newcommand{\FCalET}{\mbox{FCal $E_\mathrm{T}$}\xspace}

\newcommand{\TeV}{\mbox{Te\kern -0.1em V}\xspace}
\newcommand{\GeV}{\mbox{Ge\kern -0.1em V}\xspace}

\newcommand{\Photo}{}

\begin{document}

%\begin{frontmatter}

\title{Charged-hadron suppression in Pb+Pb and Xe+Xe collisions measured with the ATLAS detector}

%\author{Petr Balek (for the ATLAS Collaboration)\blfootnote{\textcopyright\ 2018 CERN for the benefit of the ATLAS Collaboration.}\blfootnote{Reproduction of this article or parts of it is allowed as specified in the CC-BY-4.0 license.}}
\author{Petr Balek (for the ATLAS Collaboration)}
\address{Dept. of Particle Physics and Astrophysics, Faculty of Physics, Weizmann Institute of Science, 234 Herzl Street, Rehovot 76100, Israel}

\maketitle

%% Instructions from Editor: Please use the following \dochead only in the preprint version (e-print arXiv etc.); 
%% use empty \dochead{} when submitting to Nuclear Physics A!
%\dochead{XXVIIth International Conference on Ultrarelativistic Nucleus-Nucleus Collisions\\ (Quark Matter 2018)}
% \dochead{}

\begin{abstract}

%The measurement of charged-hadron production in heavy-ion collisions provides insight into the properties of the quark-gluon plasma by measuring parton energy loss. 
The ATLAS detector at the LHC recorded 0.49\,nb$^{-1}$ of \PbPb collisions and 25\,pb$^{-1}$ of \pp collisions, both at the center-of-mass energy 5.02\,\TeV per nucleon pair. Recently, ATLAS also recorded 3\,$\mu$b$^{-1}$ of \XeXe collisions at the center-of-mass energy 5.44\,\TeV, which provides a new opportunity to study the system-size dependence of the charged-hadron production in heavy-ion collisions. The large acceptance of the ATLAS detector allows to measure the spectra of charged hadrons in a wide range of pseudorapidity and transverse momentum. The nuclear modification factors $\Raa$ are constructed as a ratio of the spectra measured in \PbPb or \XeXe collisions to that measured in \pp collisions. The $\Raa$ obtained in the two systems are presented for different centrality intervals and the results are discussed.
%\end{abstract}

\hspace*{1cm}

\noindent \textit{keywords:} xenon--xenon collisions, lead--lead collisions, charged-hadron production, nuclear modification factor
\end{abstract}

%\end{frontmatter}

%\linenumbers

\section{Introduction}

The \XeXe collisions delivered by the LHC in 2017 offer a unique opportunity to study properties of the quark-gluon plasma in systems with different geometries~\cite{Qin:2015srf,MehtarTani:2013pia}. Previous measurements~\cite{HION-2011-03,ATLAS-CONF-2017-012} show that the yields of charged hadrons are suppressed in the \PbPb collisions relative to the \pp collisions in a centrality-dependent way, when accounted for an increased parton flux in the \PbPb collisions. The new \XeXe data allow to study the centrality dependence of this suppression at a whole new angle.

The suppression of charged-hadron production is quantified using the nuclear modification factor $\Raa$:

\begin{equation}
\Raa = \frac{1}{\avgTaa} \frac{1/N_{\mathrm{evt}}\ \mathrm{d}^2 N_\mathrm{ch} /\mathrm{d}\eta\mathrm{d}\pT}{\mathrm{d}^2 \sigma_{pp}/\mathrm{d}\eta\mathrm{d}\pT},
\end{equation}
where $\avgTaa$ is the nuclear thickness function which accounts for the fact that in a nucleus--nucleus collision, a nucleon can interact with more than one nucleon from the other nucleus; $1/N_\mathrm{evt}\ \mathrm{d}^2 N_\mathrm{ch}/\mathrm{d}\eta\mathrm{d}\pT$ is the per-event yield of charged hadrons in \XeXe or \PbPb collisions measured differentially in pseudorapidity $\eta$ and transverse momentum $\pT$; and $\mathrm{d}^2 \sigma_{pp}/\mathrm{d}\eta\mathrm{d}\pT$ is the differential \pp\ cross-section.

\section{Analysis}

The measurement~\cite{ATLAS-CONF-2018-007} described in this proceeding uses \XeXe data recorded by the ATLAS detector~\cite{Aad:2008zzm} at $\sqn=5.44$\,\TeV with the total integrated luminosity of 3\,$\mu$b$^{-1}$. The \pp cross-section is obtained by extrapolation of $\sqs=5.02$\,\TeV data~\cite{ATLAS-CONF-2017-012} to the same center-of-mass energy.

The measurement is performed using the inner detector, calorimeters, muon spectrometer, trigger system and data acquisition system. The tracking information is provided by the inner detector covering \mbox{$|\eta|<2.5$}. It is immersed in a 2\,T axial magnetic field. The calorimeter system consists of an electromagnetic calorimeter covering $|\eta|<3.2$, hadronic calorimeters covering also $|\eta|<3.2$ and forward calorimeters covering \mbox{$3.1<|\eta|<4.9$.} The muon spectrometer covers $|\eta|<2.7$. 
The \XeXe events were recorded with two minimum-bias triggers. They required the total transverse energy deposited in the calorimeters to be more than 4\,\GeV or to have at least one track reconstructed in the inner detector. 

The centrality of the collisions is characterized by the total transverse energy in the forward calorimeters (\FCalET), whose distribution is divided into percentiles of the inelastic cross-section. If the nuclei overlap significantly, the collision is called ``central", while collisions with a small overlap are called ``peripheral". A~Monte Carlo Glauber model simulation~\cite{glauber,Loizides:2014vua} is used to estimate the mean
number of nucleons participating in the collision, $\avgNpart$, the mean number of binary nucleon--nucleon collisions, $\avgNcoll$, the nuclear thickness function, $\avgTaa$, as well as their uncertainties.

A particle emerging from the interaction point and passing through the inner detector typically crosses 4~layers of the pixel detector, 4~double-sided modules of the semiconductor tracker (SCT) and around 36~straw tubes of the transition radiation tracker. 
Reconstructed tracks are required to have at least 9 (11) hits, if they are within $|\eta|\leq 1.65$ ($|\eta| > 1.65$).
At least one hit is required to be in one of the two innermost layers of the pixel detector, if the tracks passed through active sensors. 
Tracks shall not have any missing hits in the pixel or SCT detectors if such hits are expected from the track trajectory.
Tracks are also required to emerge from the collisions vertex.
Tracks with $\pT>40$\,\GeV are further required to be matched to a jet within $\Delta R = \sqrt{\Delta^2 \eta + \Delta^2 \phi} < 0.4$. The jets are reconstructed in the hadronic calorimeters using the \akt algorithm~\cite{antikt} with the radius parameter of $R=0.4$. Tracks are required not to exceed the $\pT$ of the matched jets by more than 30\% in order to suppress mis-measured tracks. Such tracks are suppressed by enforcing conservation of energy, however track and jet momentum resolutions are taken into account as well.

Monte Carlo simulations are used to study the detector response effects.
Hard-scattering \pp collisions generated by \textsc{Pythia 8}~\cite{pythia8} are overlaid onto \XeXe collisions produced by \textsc{Hijing}~\cite{hijing}. The resulting events are reconstructed in the same way as data. A total of $3\cdot 10^6$ events are generated in different exclusive kinematic intervals of leading charged-hadron \pT, allowing sufficient statistics over the whole \pT range. 

There are several corrections applied to the measured spectra. 
First, leptons from the decays of electroweak bosons are subtracted as they do not follow the same suppression pattern as hadrons~\cite{ATLAS-CONF-2017-010}.
Then, secondary and fake tracks are subtracted. The former ones are tracks matched to secondary particles, and the later ones are the tracks that are coming from the spurious combination of hits not associated with a single particle. Their fraction is estimated from the simulations. It does not exceed 1\% at $\pT \approx 1$\,\GeV and is even less at $\pT \gtrsim 1$\,\GeV.
The spectra are also corrected for the $\pT$ resolution and for the track reconstruction efficiency by the bin-by-bin unfolding. The efficiency, which is also estimated from the simulations, is about 75\% at $\pT \approx 1$\,\GeV, $|\eta|\lesssim 1$ and in peripheral collisions. At $|\eta| \approx 2.5$, the efficiency decreases down to about 60\%. Another reduction, which is less than 15\%, is observed in the most central collisions. A small increase of the efficiency with increasing $\pT$ is also present, however it is no more than 5\%.

The \pp cross-section measured at $\sqs=5.02$\,\TeV is extrapolated to $\sqs=5.44$\,\TeV by the ratio of the samples generated by \textsc{Pythia 8} with $1.9\cdot 10^7$ events in each energy regime.
The ratio shows an increase of the cross-section section by about 4\% at $\pT \approx 1$\,\GeV and up to 26\% at the highest $\pT$ and $|\eta|$ measured.

There are several sources of the systematic uncertainties affecting the results. The analysis parameters are varied independently and the resulting outcomes are compared to that of the default setup. The correlated components are varied consistently in numerator and denominator in order to estimate the uncertainty on $\Raa$.
Variation of the track selection requirements introduces an uncertainty not exceeding 5\%. 
The analysis corrections depend on a matching of the reconstructed tracks to the generated particles. The uncertainty covering ambiguities in the matching procedure is about 1\%.
The bin-by-bin correction uncertainty has three sources. Limited statistics of the simulation samples yield an uncertainty of no more than 7\%. The difference of the shape of charged-hadron spectra in data and \textsc{Pythia} results in an uncertainty of 2\%. Due to the limited description of the inactive material of the detector, an uncertainty of up to 6\% has to be assigned.
An uncertainty of the geometric parameter $\avgTaa$ is largest for the peripheral collisions where it reaches about 8\%. In the central collisions, it is less than 1\%.
A half of the difference between the \pp cross-sections at $\sqs=5.02$\,\TeV and 5.44\,\TeV is assigned as a systematic uncertainty of the extrapolation.

\section{Results}
\begin{figure}[ht!]
    \centering
    \includegraphics[width=0.495\textwidth]{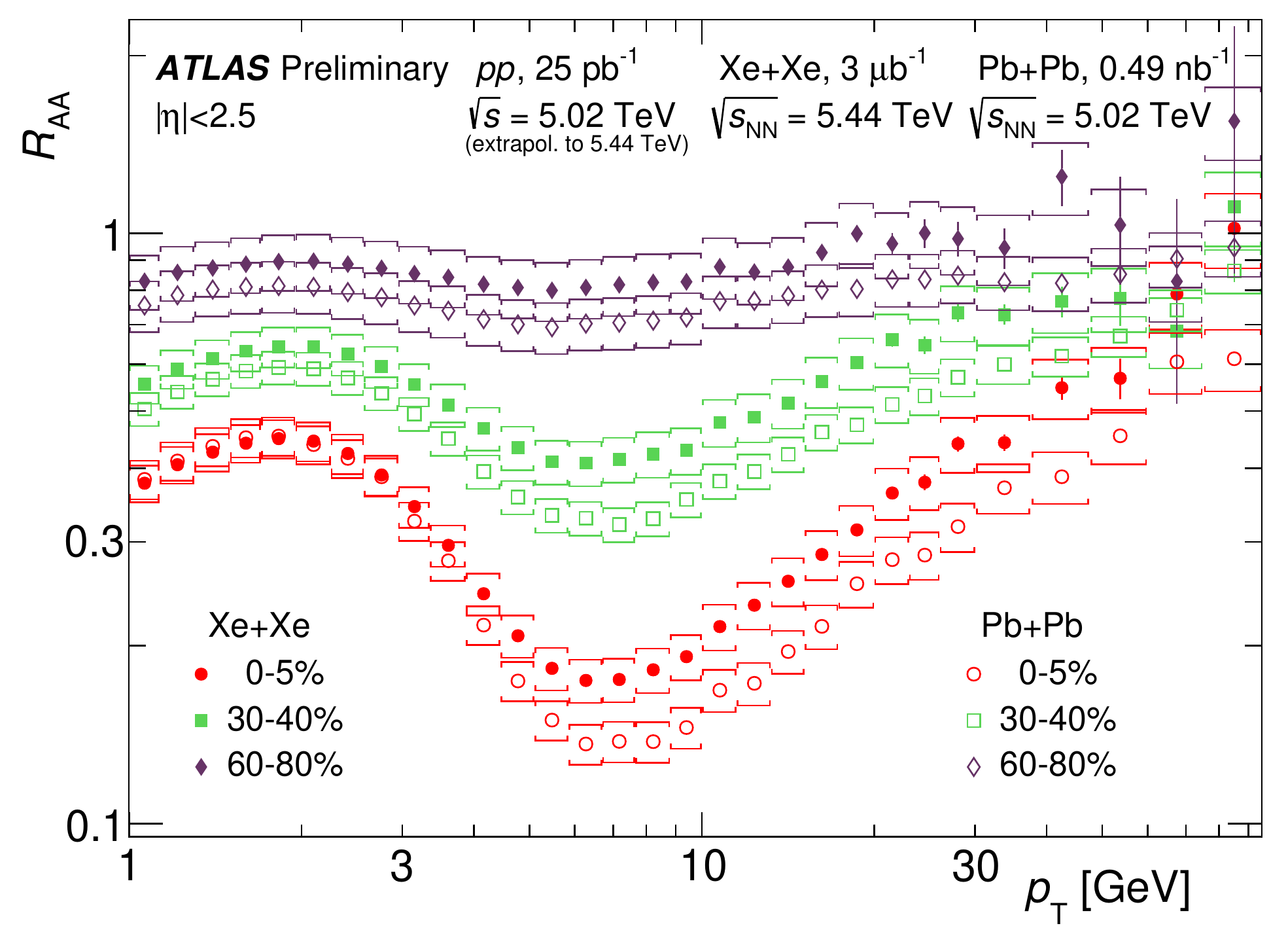} \includegraphics[width=0.495\textwidth]{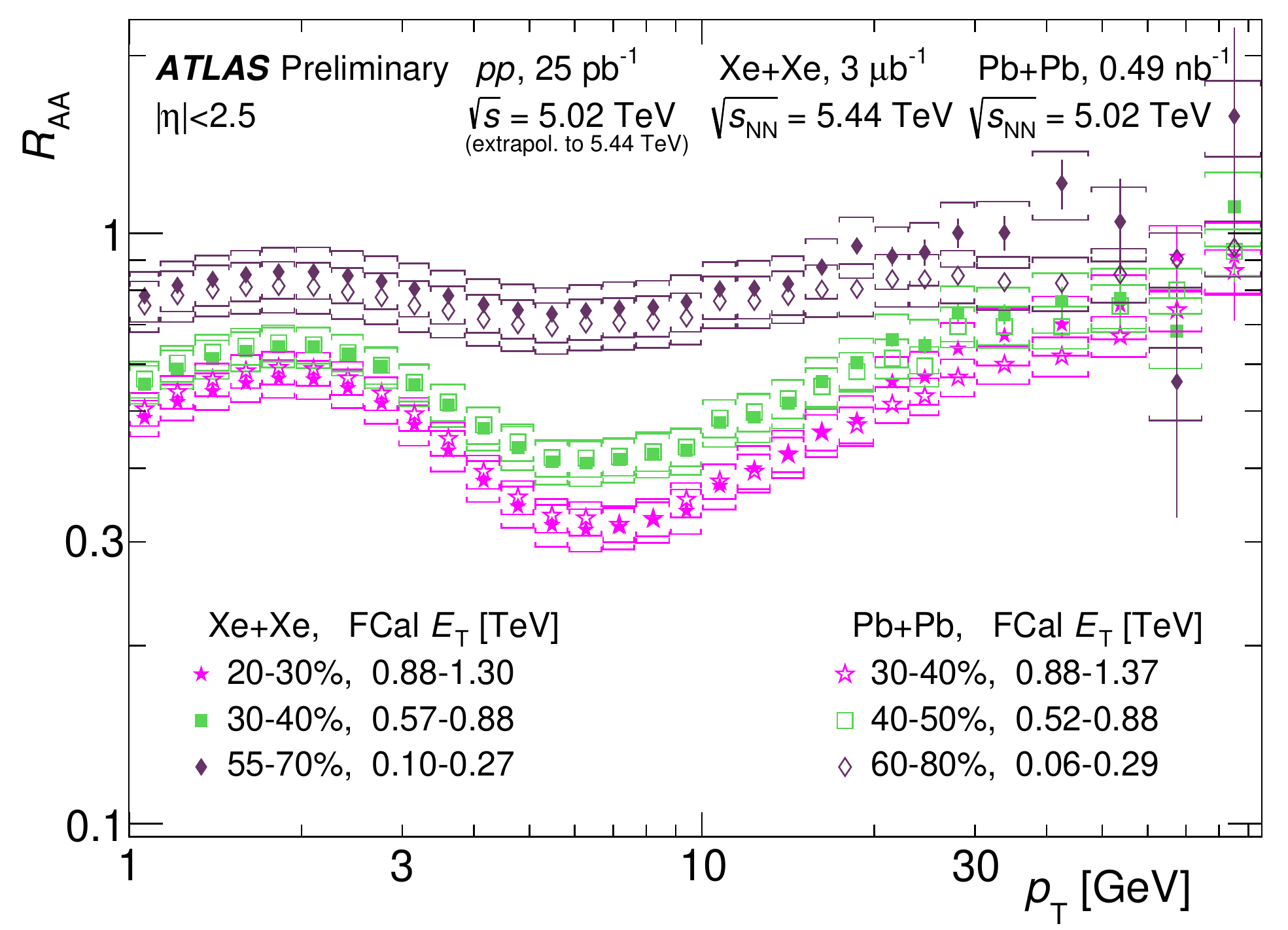}
%    \caption{Nuclear modification factors $\Raa$ as a function of \pT measured in \XeXe collisions (closed markers)~\cite{ATLAS-CONF-2018-007} and in Pb+Pb collisions (open markers)~\cite{ATLAS-CONF-2017-012}. The intervals of the same marker styles have the same centrality (left) or comparable deposited energy in the forward calorimeter (right). The statistical uncertainties are shown as the bars; systematic uncertainties are shown by the brackets.}
	\caption{Nuclear modification factors $\Raa$ as a function of \pT measured in \XeXe collisions (closed markers) and in Pb+Pb collisions (open markers). The intervals of the same marker styles have the same centrality (left) or comparable deposited energy in the forward calorimeter (right). The statistical uncertainties are shown as the bars; systematic uncertainties are shown by the brackets.}
    \label{fig1}
\end{figure}

The left panel of Fig.~\ref{fig1} 
shows the nuclear modification factors, \Raa, for \XeXe and \PbPb collisions in the same centrality intervals.
They have a characteristic curvature which is more pronounced in the central collisions. Curves reach a maximum at $\pT\approx 2$\,\GeV, then a minimum at around 7\,\GeV and then increase up to around 60\,\GeV. 
The behavior of $\Raa$ in \XeXe collisions above this value is difficult to ascertain due to the low statistics. In \PbPb collisions, the slope of $\Raa$ above $\pT \approx 60$\,\GeV diminishes. 
The stronger suppression in \PbPb than in \XeXe collisions for the same centrality intervals is expected because size of \PbPb collision system is larger than that of \XeXe collision.
\begin{figure}[ht!]
    \centering
    \includegraphics[width=0.495\textwidth]{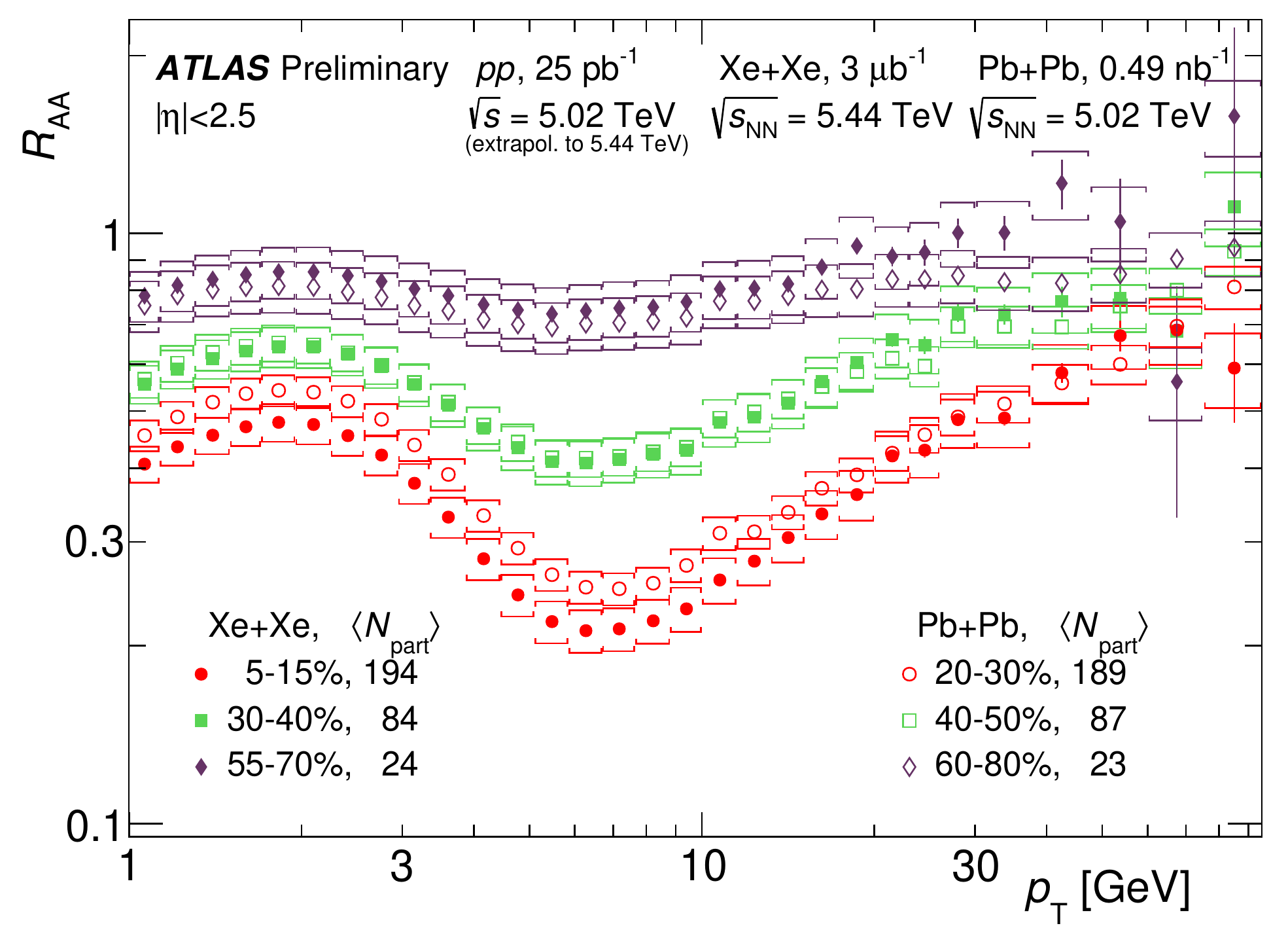}       \includegraphics[width=0.495\textwidth]{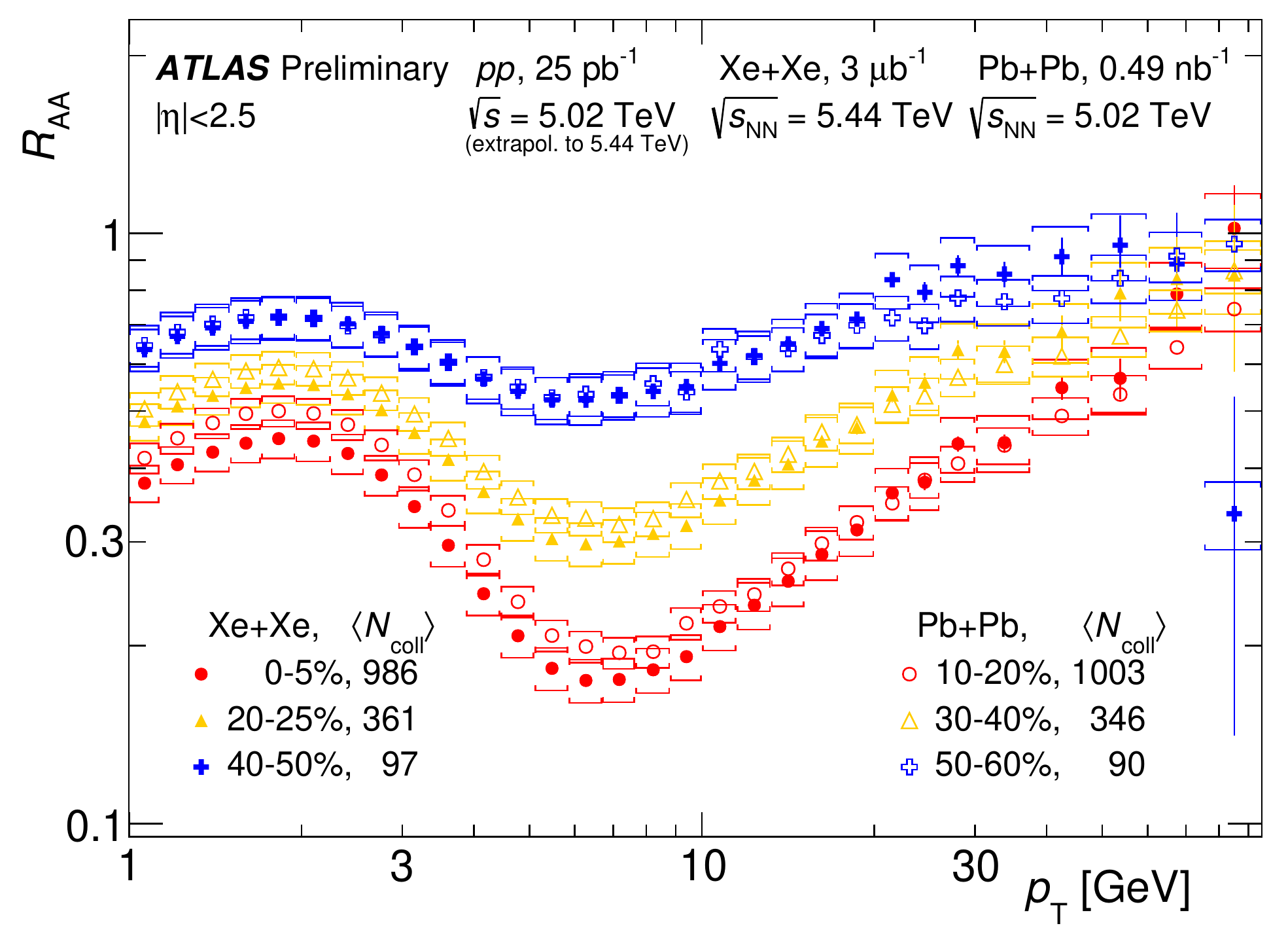}
%    \caption{Nuclear modification factors $\Raa$ as a function of \pT measured in \XeXe collisions (closed markers)~\cite{ATLAS-CONF-2018-007} and in Pb+Pb collisions (open markers)~\cite{ATLAS-CONF-2017-012}. The centrality intervals of the same marker styles have comparable $\avgNpart$ (left) or $\avgNcoll$ (right). The statistical uncertainties are shown as the bars; systematic uncertainties are shown by the brackets.}
    \caption{Nuclear modification factors $\Raa$ as a function of \pT measured in \XeXe collisions (closed markers) and in Pb+Pb collisions (open markers). The centrality intervals of the same marker styles have comparable $\avgNpart$ (left) or $\avgNcoll$ (right). The statistical uncertainties are shown as the bars; systematic uncertainties are shown by the brackets.}
    \label{fig2}
\end{figure}
The right panel of Fig.~\ref{fig1} shows $\Raa$ for \XeXe and \PbPb collisions in centrality intervals corresponding to approximately the same \FCalET. Collisions with the same \FCalET have about the same size. The observed suppressions are consistent between the two systems within the systematics uncertainties, suggesting scaling with the system size.

\begin{figure}[t!]
  \hspace*{-0.05cm}
  \begin{minipage}[c]{0.495\textwidth}
    \includegraphics[width=\textwidth]{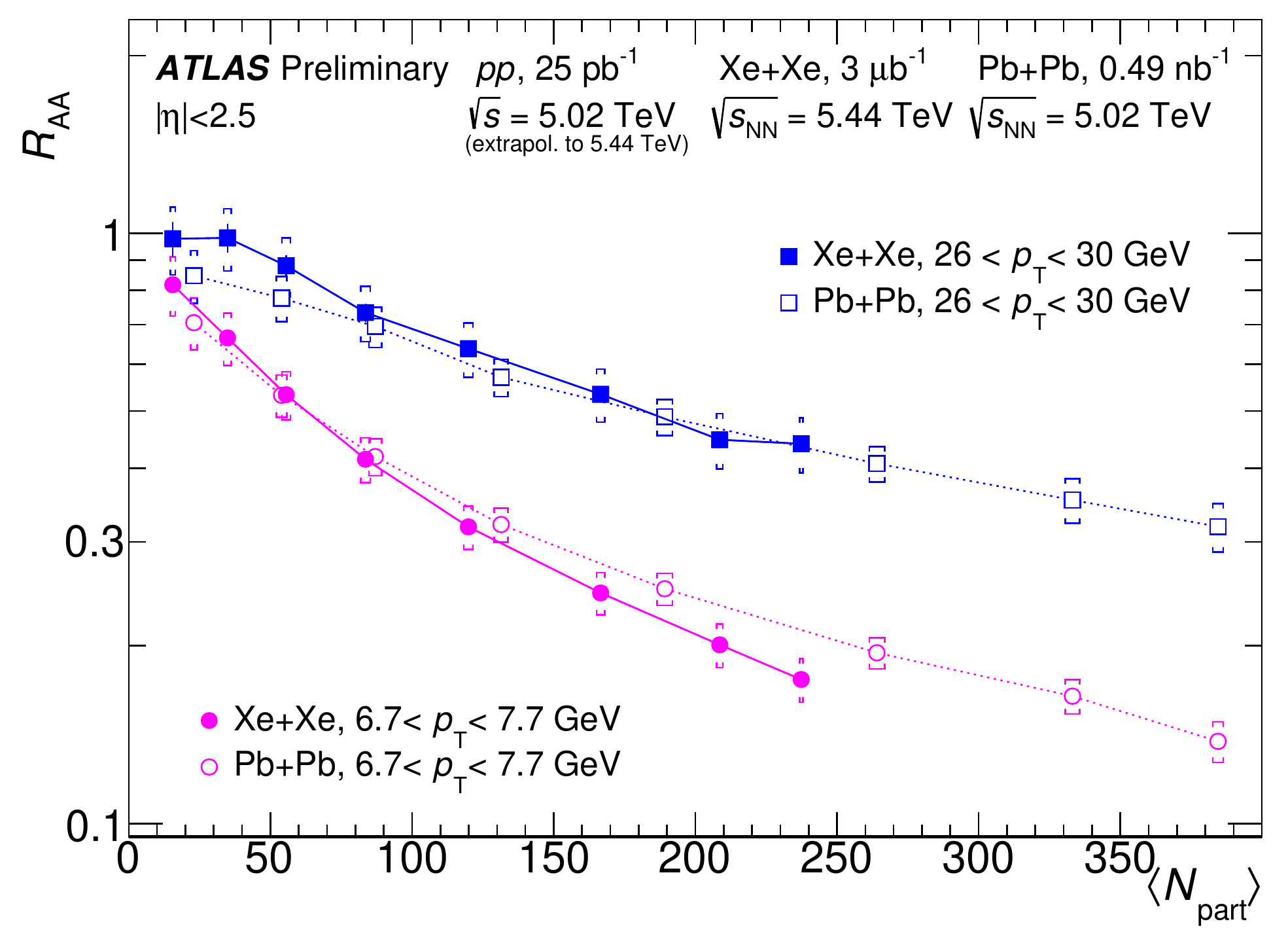}
  \end{minipage}
  \hspace*{0.63cm}
  \begin{minipage}[c]{0.35\textwidth}
%    \caption{Nuclear modification factors $\Raa$ as a function of $\Npart$ for two selected $\pT$ ranges measured in \XeXe collisions (closed markers)~\cite{ATLAS-CONF-2018-007} and in Pb+Pb collisions (open markers)~\cite{ATLAS-CONF-2017-012}. The statistical uncertainties are shown as the bars, and systematic uncertainties are shown by the brackets. The width of the brackets represents the systematic uncertainty of $\Npart$. The lines are only to help guide the eye.}
    \caption{Nuclear modification factors $\Raa$ as a function of $\Npart$ for two selected $\pT$ ranges measured in \XeXe collisions (closed markers) and in Pb+Pb collisions (open markers). The statistical uncertainties are shown as the bars, and systematic uncertainties are shown by the brackets. The width of the brackets represents the systematic uncertainty of $\Npart$. The lines are only to help guide the eye.}
    \label{fig3}
  \end{minipage}
\end{figure}

Figure~\ref{fig2} shows comparison of nuclear modification factors for \XeXe and \PbPb collisions for centrality intervals of similar $\avgNpart$ (left) and $\avgNcoll$ (right). 
The production rate of low-$\pT$ (high-$\pT$) particles is rather proportional to $\Npart$ ($\Ncoll$), and therefore the size of the two systems is expected to be comparable at similar $\Npart$. However, the agreement between the systems is still worse than in the right panel of Fig.~\ref{fig1}.  
At $\pT$ around 7\,\GeV, the \XeXe results on the left panel of Fig.~\ref{fig2} show slightly stronger suppression for the central events, but slightly milder suppression for peripheral events. This feature is demonstrated in Fig.~\ref{fig3} where it is clearly visible. 
At higher $\pT$ (26--30\,\GeV), the suppressions are comparable within uncertainties and $\Raa$ measured in both \XeXe and \PbPb collisions follow the same dependency, which suggests the suppression scales with the system size.

%\begin{figure}[th!]
%    \centering
%    \includegraphics[width=0.495\textwidth]{conf_spectra__npart}
%    \caption{Nuclear modification factors $\Raa$ as a function of $\Npart$ for two selected $\pT$ ranges measured in \XeXe collisions (closed markers)~\cite{ATLAS-CONF-2018-007} and in Pb+Pb collisions (open markers)~\cite{ATLAS-CONF-2017-012}. The statistical uncertainties are shown as the bars, and systematic uncertainties are shown by the brackets. The width of the brackets represents the systematic uncertainty of $\Npart$. The lines are only to help guide the eye.}
%    \label{fig3}
%\end{figure}

\section{Summary \& Acknowledgements}

Measurement of charged-hadron spectra and the nuclear modification factor in \XeXe collisions has been presented. The $\Raa$ is compared between the \XeXe collisions at $\sqn=5.44$\,\TeV and \PbPb collisions at $\sqn=5.02$\,\TeV measured by the ATLAS detector at the LHC. 

The data suggest that $\Raa$ scales with the system size. Other aspects of the collisions, such as center-of-mass energy or initial energy density, may not play a significant role for the comparison presented in this proceedings.
However, they may become important when comparing collisions at the LHC energies to those at e.g.~RHIC energies. 

%\section*{Acknowledgements}
This research is supported by the Israel Science Foundation (grant 1065/15) and by the MINERVA Stiftung with the funds from the BMBF of the Federal Republic of Germany.

\bibliographystyle{elsarticle-num}
\bibliography{nupha-PetrBalek-QM18}

\begin{thebibliography}{10}
\expandafter\ifx\csname url\endcsname\relax
  \def\url#1{\texttt{#1}}\fi
\expandafter\ifx\csname urlprefix\endcsname\relax\def\urlprefix{URL }\fi
\expandafter\ifx\csname href\endcsname\relax
  \def\href#1#2{#2} \def\path#1{#1}\fi

\bibitem{Qin:2015srf}
G.-Y. Qin, X.-N. Wang, Int. J. Mod. Phys. E 24~(11) (2015) 1530014.
\newblock \href {http://arxiv.org/abs/1511.00790} {\path{arXiv:1511.00790}}.

\bibitem{MehtarTani:2013pia}
Y.~Mehtar-Tani, J.~G. Milhano, K.~Tywoniuk, Int. J. Mod. Phys. A 28 (2013)
  1340013.
\newblock \href {http://arxiv.org/abs/1302.2579} {\path{arXiv:1302.2579}}.

\bibitem{HION-2011-03}
{ATLAS Collaboration}, JHEP 09 (2015) 050.
\newblock \href {http://arxiv.org/abs/1504.04337} {\path{arXiv:1504.04337}}.

\bibitem{ATLAS-CONF-2017-012}
{ATLAS Collaboration}, {ATLAS-CONF-2017-012},
  \url{https://cds.cern.ch/record/2244824}.

\bibitem{ATLAS-CONF-2018-007}
{ATLAS Collaboration}, ATLAS-CONF-2018-007.
  \url{https://cds.cern.ch/record/2318588}.

\bibitem{Aad:2008zzm}
{ATLAS Collaboration}, JINST 3 (2008) S08003.

\bibitem{glauber}
M.~L. Miller, et~al., Ann. Rev. Nucl. Part. Sci. 57 (2007) 205--243.
\newblock \href {http://arxiv.org/abs/nucl-ex/0701025}
  {\path{arXiv:nucl-ex/0701025}}.

\bibitem{Loizides:2014vua}
C.~Loizides, J.~Nagle, P.~Steinberg, SoftwareX 1--2 (2015) 13--18.
\newblock \href {http://arxiv.org/abs/1408.2549} {\path{arXiv:1408.2549}}.

\bibitem{antikt}
M.~Cacciari, G.~P. Salam, G.~Soyez, JHEP 04 (2008) 063.
\newblock \href {http://arxiv.org/abs/0802.1189} {\path{arXiv:0802.1189}}.

\bibitem{pythia8}
T.~Sj{\"o}strand, et~al., Comput. Phys. Commun. 191 (2015) 159--177.
\newblock \href {http://arxiv.org/abs/1410.3012} {\path{arXiv:1410.3012}}.

\bibitem{hijing}
X.-N. Wang, M.~Gyulassy, Phys. Rev. D 44 (1991) 3501--3516.

\bibitem{ATLAS-CONF-2017-010}
{ATLAS Collaboration}, ATLAS-CONF-2017-010.
  \url{https://cds.cern.ch/record/2244821}.

\end{thebibliography}

\end{document}